\documentclass[journal,12pt,onecolumn,draftclsnofoot]{IEEEtran}
\usepackage[top=0.75in,bottom=0.75in,right=0.75in,left=0.75in]{geometry}
\makeatletter
\def\ps@headings{%
\def\@oddhead{\mbox{}\scriptsize\rightmark \hfil \thepage}%
\def\@ adversarynhead{\scriptsize\thepage \hfil \leftmark\mbox{}}%
\def\@oddfoot{}%
\def\@ adversarynfoot{}}
\makeatother
\ifCLASSINFOpdf
\else
 
\fi
 
\usepackage{epsfig}
\usepackage{array}
\newcolumntype{L}[1]{>{\raggedright\let\newline\\\arraybackslash\hspace{0pt}}m{#1}}
\newcolumntype{C}[1]{>{\centering\let\newline\\\arraybackslash\hspace{0pt}}m{#1}}
\newcolumntype{R}[1]{>{\raggedleft\let\newline\\\arraybackslash\hspace{0pt}}m{#1}}
\usepackage{amsmath}
\usepackage[short]{optidef}
\usepackage{amsthm} 
\usepackage{mathrsfs}
\newcommand{\bc}{\begin{center}}
\newcommand{\ec}{\end{center}}
\newcommand{\be}{\begin{equation}}
\newcommand{\ee}{\end{equation}}
\usepackage{flexisym}
\usepackage{algorithmicx}
\usepackage{algorithm}
\usepackage{soul}
\usepackage{algpseudocode}

\usepackage{amsmath}

\newcommand{\bnu}{\begin{enumerate}}
\newcommand{\enu}{\end{enumerate}}
\usepackage{cite}
\usepackage{url}
\usepackage{breqn}
\usepackage{lipsum}
\usepackage{mathtools}
\usepackage{amsmath}
\usepackage{caption}
\usepackage{subcaption}
\usepackage{soul}
\usepackage{blindtext, graphicx}
\usepackage{cuted}
\usepackage{color}

\usepackage{amsfonts}
\usepackage{lipsum}
\usepackage{cuted}
\newtheoremstyle{case}{}{}{}{}{}{:}{ }{}

\begin{document}
\newgeometry{top=1in,bottom=0.95in,right=0.75in,left=0.75in}
\title{Blockage Prediction for Mobile UE in RIS-assisted Wireless Networks: A Deep Learning Approach}
\author{Shakil Ahmed,~\IEEEmembership{Student Member,~IEEE}, Ibrahim Abdelmawla, Ahmed E. Kamal,~\IEEEmembership{Fellow,~IEEE}, and Mohamed Y. Selim,~\IEEEmembership{Senior Member,~IEEE}
\vspace*{-1.25cm}
\thanks{ Shakil Ahmed, Ibrahim Abdelmawla, Ahmed E. Kamal, and Mohamed Y. Selim are with the Department of Electrical and Computer Engineering, Iowa State University, Ames, Iowa, USA. (email: \{shakil, ibrahim4, kamal, myoussef\}@iastate.edu).
}}
 

\maketitle

\IEEEpeerreviewmaketitle

\begin{abstract}
Due to significant blockage conditions in wireless networks, transmitted signals may considerably degrade before reaching the receiver. The reliability of the transmitted signals, therefore, may be critically problematic due to blockages between the communicating nodes. Thanks to the ability of Reconfigurable Intelligent Surfaces (RISs) to reflect the incident signals with different reflection angles, this may counter the blockage effect by optimally reflecting the transmit signals to receiving nodes, hence, improving the wireless network's performance. With this motivation, this paper formulates a RIS-aided wireless communication problem from a base station (BS) to a mobile user equipment (UE). The BS is equipped with an RGB camera.
We use the RGB camera at the BS and the RIS panel to improve the system's performance while considering signal propagating through multiple paths and the Doppler spread for the mobile UE.
First, the RGB camera is used to detect the presence of the UE with no blockage.
When unsuccessful, the RIS-assisted gain takes over and is then used to detect if the UE is either ``present but blocked'' or ``absent''.
The problem is determined as a ternary classification problem with the goal of maximizing the probability of UE communication blockage detection. We find the optimal solution for the probability of predicting the blockage status for a given RGB image and RIS-assisted data rate using a deep neural learning model. We employ the residual network 18-layer neural network model to find this optimal probability of blockage prediction. Extensive simulation results reveal that our proposed RIS panel-assisted model enhances the accuracy of maximization of the blockage prediction probability problem by over 38\% compared to the baseline scheme.
\\ \indent {\em Keywords---}{\bf RGB camera, RIS panel, Doppler spread, user equipment, deep neural network, blockage prediction, residual network-18 layer.}\\
\end{abstract}
\vspace*{-.75cm}

\section{Introduction}
Recently, Reconfigurable Intelligent Surfaces (RISs) have emerged as a potential solution to tackle various critical challenges in the wireless domain, such as blockage mitigation, resource allocation, capacity improvement, etc. \cite{Ref_1,Ref_2}.
Obstacles blocking the wireless links between the transmitter and receiver always raise serious concern that hinders maximizing the system performance with optimal quality of service.
Fortunately, the RIS panel mitigates this concern by providing improved benefits, configuring the wireless propagation compared to conventional relaying systems, such as amplify-and-forward, while minimizing power consumption and blockage effect.
The RIS panel deployment creates a software-defined intelligent environment, hence, facilitating an enhanced signal-to-noise ratio (SNR) for various areas, including urban areas with obstacles.
Moreover, the RIS panel has minimal hardware complexity \cite{Ref_4}.
 
Increasing the number of antenna elements at the transmitter and receiver side often results in hardware complexity, maintenance, and cost overhead.
As future wireless communications move to higher service quality, the channel propagation characteristics must be modified extensively to meet the need.
A major and critical challenge is to maintain the stability of signals, which can be done by leveraging the ability to reflect the transmit signals from the surfaces and, therefore, minimizing the power budget during the communications process.
This results in a strong emphasis on the demand for antenna directives to achieve higher SNR.
Hence, blockage prediction becomes essential for next-generation wireless technology systems to attain such a goal.
Unfortunately, the probability of the blockage prediction requires the challenging task of attaining accurate and efficient classical signal processing \cite{Ref_6}.
The challenge becomes even more critical for mobile nodes.
These requirements may result in the challenge of accommodating the reliability of mobility-based wireless systems.
 
\subsection{Background}
\textbf{What is an RIS panel?}
The RIS panel is made of electromagnetic metasurface material, which may be deployed on indoor and outdoor objects to attain improved channel gains without compromising the standardization and hardware flexibility of traditional next-generation technology.
The RIS panel is made of a number of passive reflective elements, which are intelligently programmed to control the propagation of wireless signals, eventually creating virtual and stable line-of-sight links among the communicating nodes, such as between a base station (BS) and the user equipment (UE).
The passive elements of the RIS panels have reconfigurable parameters, such as phase shifts and attenuation \cite{Ref_7}.
The authors in \cite{Ref_8, Ref_9} designed the reflection arrays of an RIS panel that controls the phases with low computational complexity.
Several approaches were proposed to design the RIS panel phase shift \cite{Ref_10}.

\textbf{Machine learning for wireless communications:}
Wireless communications system performance has been enhanced by using machine learning tools, such as deep learning, by addressing challenges, such as blockage prediction in various domains, massive channel prediction \cite{Ref_11}, proactive hand-off \cite{Ref_12}, and intelligent resource allocations \cite{Ref_13}.
It is necessary to use efficient datasets to advance research on machine learning in the management of vision and RIS panel-assisted wireless transmission.
Such a dataset is publicly available, called ViWi \cite{Ref_ViWi}.
The authors \cite{MA_1} investigated a deep learning-based model, where the BS was equipped with an RGB camera to select the best beam in sensitive blockage scenarios to exploit the potential benefits of wireless systems.
All the prior work in \cite{Ref_4}~-~\cite{MA_1} assumed that the channel state information between the transmitter and the receiver might have been perfect or partially known with considerable error.
However, acquiring this channel state information with tolerable error is the most critical parameter for wireless systems since the next-generation technology uses massive numbers of antennas with physical constraints.
The Doppler effect due to the UE mobility must be taken into account for practical consideration of wireless networks.
Moreover, adding the RIS panel to the wireless network further enhances the system performance.
\vspace*{-0.40 cm}
\subsection{Contributions}
The contributions of this paper can be summarized as follows:
We investigate the issue of blockage prediction for various locations of a mobile UE in the RIS-assisted wireless communication environment.
We adopt the Doppler spread to capture the effect of the mobility of the UE on the channel characteristics.
At first, the RGB camera is used to detect the presence of UE.
This paper also considers deploying an RIS panel to differentiate between an absent or present but unblocked UE once the RGB camera cannot detect the unblocked UE.
The unsuccessful detection of a UE, therefore, can be confirmed or mitigated using the RIS panel, as shown in Fig.~1.
This allows detecting an absent, blocked, and unblocked UE at various locations using the RGB camera and RIS panel.
We propose a deep learning-based framework to train the model to maximize the blockage probability prediction using an RGB camera at the BS and an RIS panel in the wireless channel.
Note that we use the dataset from ViWi \cite{Ref_ViWi} to capture the RGB images from the BS.
Detecting a blockage using the RGB camera, hence using the image, is a bit more difficult as the cases of absent UE and blocked UE are visually identical.
This UE location detection is investigated as a ternary classification problem.
Therefore, the RGB images and RIS panel-assisted-channel gains are coupled to detect the status of UE.
In short, the proposed system model maximizes the probability of blockage prediction for various locations of UE with the help of the RIS panel and ViWi dataset.

 
\section{System Model} \label{SystemModel}
\begin{figure}[!h]
\centering
\includegraphics[width=4.80in]{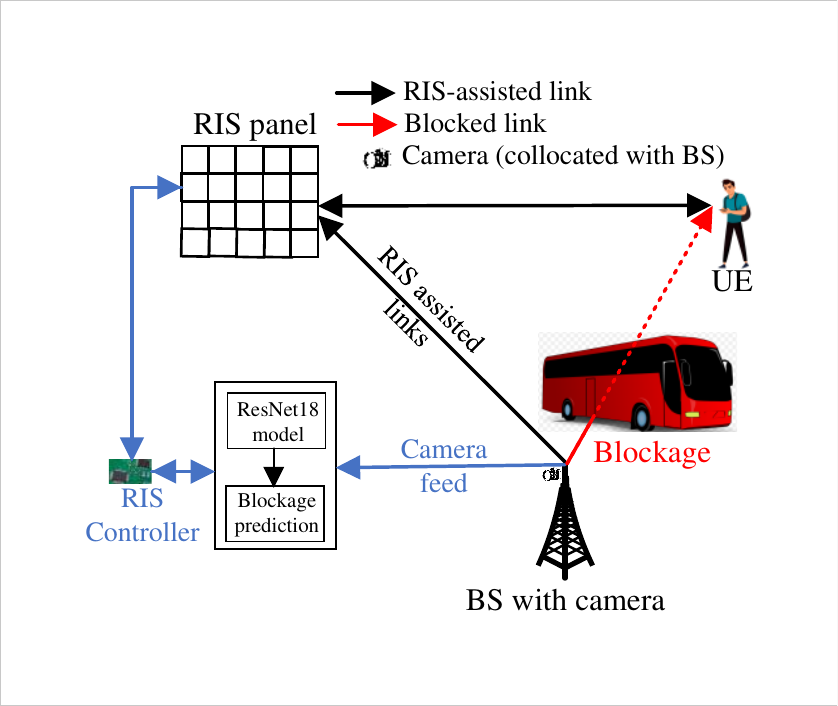}
\caption{Blockage prediction model}
\label{System_Model2}
\end{figure}
As shown in Fig.~\ref{System_Model2}, we consider a downlink scenario of the RIS panel-assisted wireless network, where a BS communicates with a mobile UE.
The BS, equipped with an RGB camera, has a uniform linear array antenna pattern with $M$ antenna elements, while the UE is equipped with a single antenna.
For practicality, the BS employs an analog architecture and contains a single RF chain.
The RIS panel has $R$ passive elements, and we assume the passive elements are ideal and control the incident signals' reflection angle and phase shift.

\subsection{Properties of RIS Panel}
The RIS passive elements reflect the incident signals towards the destination of interest.
Each passive element is smaller than the wavelength.
The elements adjust the reflection coefficient to re-experience the signals with nearly the same gain and proper beamforming into the directions of interest.
The RIS panel elements properties are expressed by a diagonal matrix $R \times R$:
\be
\begin{aligned}
\label{Delta}
\boldsymbol{\Delta} = \mathrm {diag} (\alpha_1 e^{j \delta_{1}}, \alpha_2 e^{j \delta_2}, ..., \delta_{R} e^{j \delta_{R}})
\end{aligned}
\ee
where $\alpha_i \in [0,1] $ defines amplitude reflection coefficient and $\theta_i$ is the passive element phase-shift, both for element $i$.
\begin{table}[]
\caption{List of symbols used in the paper}
\begin{center}
\begin{tabular}{|c|l|c|l|}
\hline
Symbol & Description & Symbol & Description \\ \hline
$f_s$ & Doppler spread & $X$ & Input image \\ \hline
$R$ & RIS elements & $H,W,C$ & Height, width, color \\ \hline
$\lambda$ & Wavelength & $k^{b}$ & BS-UE $u$ MPCs\\ \hline
$\boldsymbol{\Delta}$ & RIS phase shift & $f$ & Carrier frequency \\ \hline
$U$ & UE locations & $\mathbf{H_u^o}$ & Gain for $u$ locations \\ \hline
$K^b$ & BS-UE $u$ MPCs & $K^r$ & BS-RIS $u$ MPCs \\ \hline
$d_k^b$ & Cyclic prefix of $k^{b}$ & $d_k^r$ & Cyclic prefix of $k^{r}$ \\ \hline
$\alpha_k^b$ & Path gain of $k^{b}$ & $\alpha_k^r$ & Path gain of $k^{r}$ \\ \hline
$\tau_k^b$ & Delay of $k^{b}$ & $\tau_k^r$ & Delay of $k^{r}$ \\ \hline
$\tau_k^b$ & Delay of $k^{b}$ & $\tau_k^r$ & Delay of $k^{r}$ \\ \hline
$t_k^b$ & Sampling time of $k^{b}$ & $t_k^r$ & Sampling time of $k^{r}$\\ \hline
$\theta_k^b$ & Azimuth angle of $k^{b}$ & $\theta_k^r$ & Azimuth angle of $k^{r}$ \\ \hline
$\phi_k^{b}$ & Elevation angle of $k^{b}$ & $\phi_k^{r}$ & Elevation angle of $k^{r}$ \\ \hline
$K^u$ & RIS-UE $u$ MPCs & $c$ & Speed of light \\ \hline
$d_k^u$ & Cyclic prefix of $k^{u}$ & $v$ & Speed of UE \\ \hline
$\alpha_k^u$ & Path gain of $k^{u}$ & $\delta_i$ & Element phase shift \\ \hline
$\tau_k^u$ & Delay of $k^{u}$ & $\alpha_i$ & Reflection coefficient \\ \hline
$\tau_k^u$ & Delay of $k^{u}$ & $\mathbf{h_{B}}$ & BS-to-UE $u$ link \\ \hline
$t_k^u$ & Sampling time of $k^{u}$ & $\mathbf{h_{R}}$ & BS-to-RIS link \\ \hline
$\theta_k^u$ & Azimuth angle of $k^{u}$ & $\mathbf{h_{u}}$ & RIS-to-UE $u$ link \\ \hline
$\phi_k^{u}$ & Elevation angle of $k^{u}$ & $\mathbf{I}$ & Identity matrix \\ \hline
$Q_{\Gamma}$ & Prediction function & $b_i$ & Probability \\ \hline
$o$ & Cross entropy loss & $l_u$ & Link status \\ \hline
\end{tabular}
\end{center}
\end{table}
 
\subsection{Channel Modeling }
These characteristics include the dependence on the environment geometry, frequency band, etc. As shown in Fig.~\ref{System_Model2}, the UE receives the transmitted signals through the RIS panel.
With the change of the UE locations, the blockage scenario changes.
Hence, predicting the blockage is crucial due to the mobility of the UE to retain the system's required quality of service.
In wireless networks, many MPCs may be clustered into one or more spatial lobes, each represented by a mean angle of arrival (AoA).
The authors in \cite{MKS_1} reported that the received beam was contained within a single spatial lobe.
Therefore, the proposed system model safely assumes that the received beam falls within one spatial lobe between two nodes while the other lobes have a trivial effect on the system.
We assume the UE moves between $U$ discrete locations while it travels, where $u=1,2,...,U$ and $u$ is the index that refers to the current UE location.
We define the number of MPCs as
$K=K^{b}+K^{r}+K^{u}$, where $K^{b}$, $K^{r}$, and $K^{u}$ are the total number of resolvable MPCs during BS to UE location $u$, BS to RIS, and RIS to UE location $u$ links, respectively.

\subsubsection{Channel modeling within BS and UE}
The channel gain from the BS to UE location $u$ is expressed as follows \cite{MA_1}:
\be
\begin{aligned}
\mathbf{{h}_{B}}[k^{b}]\!\!=\!\sum_{k^{b}=1}^{K^{b}}\sum_{d_k^{b}=0}^{D_k^{b}-1}
\! \! \alpha_k^{b} e^{-\mathrm{j}\frac{k^{b}}{K^{b}}\Phi_k^{b}}\!
(d_k^{b}t_k^{b}\!- \tau_k^{b}) \mathbf{a}(\theta_k^{b}, \phi_k^{b})
\end{aligned}
\ee
where $\mathbf {h_{B}} \in \mathbb{C}^{1 \times M}$.
$K^{b}$ is the number of resolvable MPCs between the BS and UE location $u$ communication, and $k^b=1,2,...,K^b$.
$d_k^{b}$ is the cyclic prefix of the $k^{b}$ MPC, where $d_k^b=1,2,...,D_K^b$ and $D_k^b$ is the total number of cyclic prefix.
$\alpha_k^{b}$ is the path gain of $k^{b}$ MPC.
$t_k^{b}$ is the sampling time of the $k^{b}$ MPC.
$\tau_k^{b}$ is the delay of $k^{b}$ MPC.
$\theta_k^{b}$, $\phi_k^{b}$ are the azimuth and elevation AoA, respectively, of the $k^{b}$ MPC.
$\mathbf{a}(\theta_k^{b}, \phi_k^{b}) \in \mathbb{C}^{1 \times M}$ defines the vector of AoA to capture the MPCs within AoA.
$\Phi_k^{b}$ is defined as follows:
\begin{align}
\Phi_k^{b}=2 \pi f \tau_k^{b} -2 \pi f_s t_k^{b} \cos(\theta_k^{b})-\phi_k^{b}
\end{align}
where $f$ is the carrier frequency.
$f_s=\frac{fv}{c}$ is the Doppler spread \cite{DS}, $v$ is the UE speed, and $c$ is the speed of light.
 
\subsubsection{Channel modeling between BS and RIS}
The channel gain from the BS to RIS is expressed as follows:
\be
\begin{aligned}
\mathbf{{h}_{R}}[k^{r}]=\sum_{k^{r}=1}^{K^{r}} \sum_{d_k^{r}=0}^{D_k^{r}-1} \alpha_k^{r} e^{-\mathrm{j}\frac{k^{r}}{K^{r}}\Phi_k^{r}}(d_k^{r}t_k^{r}- \tau_k^{r}) \mathbf{b}(\theta_k^{r}, \phi_k^{r})
\end{aligned}
\ee
where $\mathbf {h_{R}} \in \mathbb{C}^{R \times M}$.
$K^{r}$ is the number of resolvable MPCs between the BS and the RIS panel, where $k^r=1,2,...,K^r$.
$D_k^{r}$, $\alpha_k^{r}$, $t_k^{r}$, $\tau_k^r$, $\theta_k^{r}$, $\phi_k^{r}$, and $\mathbf{b}(\theta_k^{r}, \phi_k^{r})$ are the cyclic prefix, the path gains, the sampling time, the delay, the azimuth angle, elevation angle, and the matrix of AoA to capture the MPCs within mean AoA, respectively, between the BS and the RIS panel.
$\mathbf{b}(\theta_k^{r}, \phi_k^{r}) \in \mathbb{C}^{R \times M}$
$\Phi_k^{r}$ is defined as follows:
\begin{align}
\label{Eq_noDS}
\Phi_k^{r}=2 \pi f \tau_k^{r} -2 \pi t_k^{r} \cos(\theta_k^{r})-\phi_k^{r}
\end{align}
 
Note that the BS and the RIS panel are static. Hence, (\ref{Eq_noDS}) safely ignores the effect of the Doppler spread.
 
\subsubsection{Channel modeling within RIS and UE}
The channel gain from the RIS to UE location $u$ is expressed as follows:
\be
\begin{aligned}
\mathbf{{h}_{u}}[k^{u}]=\sum_{k^{u}=1}^{K^{u}} \sum_{d_k^{u}=0}^{D_k^{u}-1} \alpha_k^{u} e^{-\mathrm{j}\frac{k^{u}}{K^{u}}\Phi_k^{u}}(d_k^{u}t_k^{u}- \tau_k^{u}) \mathbf{c}(\theta_k^{u}, \phi_k^{u})
\end{aligned}
\ee
where $\mathbf {h_{R}} \in \mathbb{C}^{1 \times R}$.
$K^{u}$ is the number of resolvable MPCs between the RIS panel and UE, where $k^u=1,2,...,K^u$.
$D_k^{u}$, $\alpha_k^{u}$, $t_k^{u}$, $\tau_k^u$, $\theta_k^{u}$, $\phi_k^{u}$, and $\mathbf{c}(\theta_k^{u}, \phi_k^{u})$ are the cyclic prefix, the path gains, the sampling time, the delay, the azimuth angle, elevation angle, and the vector of AoA to capture the MPCs within mean AoA, respectively, between the RIS and the UE location $u$.
$\mathbf{c}(\theta_k^{u}, \phi_k^{u})\in \mathbb{C}^{1 \times R}$.
$\Phi_k^{u}$ is defined as follows:
\begin{align}
\Phi_k^{u}=2 \pi f \tau_k^{u} -2 \pi f_st_k^{u} \cos(\theta_k^{u})-\phi_k^{u}
\end{align}
 
\subsubsection{Overall channel gain}
For any given RIS panel element phase shift matrix, $\boldsymbol{\Delta}$, the effective channel gain from the BS to UE location $u$ is:
\be
\begin{aligned}
\label{OH}
\mathbf {H_{u}^o}[k] = \mathbf {h_{B}}[k^{b}] + \mathbf {h_{u}}[k^{u}] \mathbf {\Delta} \mathbf {h_{R}}[k^{r}]
\end{aligned}
\ee
where $k=k^b+k^r+k^u$, which is the resolvable MPCs coming through the RIS panel to UE location $u$.
$\boldsymbol{\Delta} $ is defined in (\ref{Delta}).
$\mathbf{H_u^o}[k]$ is the gain through the RIS panel to UE location $u$.
 
\subsection{Data Rate}
The rate between the BS and the UE location $u$ through RIS panel is $\mathbf{R_{u}}[k]=\log(\mathbf{I}+\mathbf{H_{u}^o}[k])$.

\subsection{RGB Camera}
The RGB camera at the BS is equipped with complementary metal-oxide-semiconductor sensors and acquires colored images of UE and obstacles.
The RGB camera computes the pixels based on their colors.
The input image, $X$ has the dimensions of $H \times W \times C$, where $H$, $W$, and $C$ define the height, width, and the number of colors of $X$, respectively.
Eventually, the RGB camera combines the color image captured directly through its filter with the other two colors captured by the pixels around it.

\section{Blockage Prediction}
 
\subsection{Problem Formulation}
We determine the blocked and unblocked UE's connections for location $u$ with the help of RGB image and RIS panel to increase the reliability of wireless communications.
We introduce an ordered pair $(X,\log(\mathbf{I}+\mathbf{H_u^o}[k]))$, where the RGB images is $X$ and the UE data RIS panel assisted rate is $\log(\mathbf{I}+\mathbf{H_u^o}[k])$.
The link status of UE at location $u$ is expressed as $l_u \in \{ -1,0, 1\}$, where $-1$ defines the absent UE, $0$ is present and unblocked UE, and $1$ defines present but blocked UE.
Maximizing the probability of blockage prediction over a number of UE locations $U$ is:
\be
\begin{aligned}
\label{Eq_opt}
\! \!\! \!\max _{Q_{\Gamma}(X, \log(\mathbf{I}+\mathbf{H_u^o}[k]))} \prod_{u=1}^{U} \mathbb{P}\! \left(\hat{l}_{u}=l_{u} \! \mid \! \left(X, \log(1\! +\! \mathbf{H_u^o}[k]) \right)\right)
\end{aligned}
\ee
where $u=1,2,...,U$ is the various locations of the mobile UE.
The blockage prediction quantifies the degree of blockage incurs during the mobility for $U$ locations. We, therefore, maximize the probability of predicted blockage using $(X,\log(\mathbf{I}+\mathbf{H_u^o}[k]))$.
Note that in (\ref {Eq_opt}), the product of the UE link status at location $u$ is conditionally independent.
$Q_{\Gamma}\left(X,\log(\mathbf{I}+\mathbf{H_u^o}[k])\right)$ is expressed as a prediction function and parameterized by a set of parameters $\Gamma$.
The output of the prediction function is a probability distribution function, $
\cal B$, of the UE locations over $U$, where $\cal B$ $= \{ b_1,b_2,...,b_U\}$.
We define the prediction function as follows:

\textbf{Definition}
\textit{Recall that $Q_{\Gamma }(X,\log(\mathbf{I}+\mathbf{H_u^o}[k]))$ function is parameterized by a set $\Gamma$ representing the parameters at the input.
Let $L=(X,\log(\mathbf{I}+\mathbf{H_u^o}[k]))$, and
$\mathbb{P}(L,l)$ represents a joint probability distribution governing the relation between the observed sequence of RGB images and RIS-assisted link to detect the blockage, $L$.
Whether blocked or not, the future location-independent link status is defined as $l$, reflecting the probabilistic nature of link blockages for a $U$ number of UE locations.
An independent location $u$ of $\mathbb{P}(L_u, l_u)$ is sampled from $\mathbb{P}(L, l)$ serves as a label for the observed $L_u$. This sampling helps maximizing the prediction function $Q_{\Gamma }(X,\log(\mathbf{I}+\mathbf{H_u^o}[k]))$ to maintain high-fidelity predictions.}

The blockage probability index, $i$, of the UE locations over $U$ are also independent.
We express $i$ in the following formulation as follows:
\begin{align}
i=\arg \! \max_{i \in \{1,2,...,U \}} \{ b_1,b_2,....,b_i,...b_U\}
\end{align}
where $b_i$ defines the blockage probability for the UE location $i$.
Neural network model maximizes the prediction function,
$Q_\Gamma \left(X,\log(\mathbf{I}+\mathbf{H_u^o}[k])\right)$, and then predicts the probability of the blockage for UE location $u$, using $Q_\Gamma \left(X,\log(\mathbf{I}+\mathbf{H_u^o}[k])\right)$ (details are provided in Section~III-B).
$Q_{\Gamma}\left(X,\log(\mathbf{I}+\mathbf{H_u^o}[k])\right)$ is, therefore, computed in a way to find the probability mass function of $U$, hence, maximizing the probability of blockage prediction.
In other words, while the objective function maximizes the probability of blockage prediction, the model finds the optimal prediction function for a given RGB image and RIS panel-assisted link, as well as $\Gamma$.
The image of the RGB camera and RIS-assisted link are fed to the neural network model to determine the status of the UE at $u$ location.
We have three outcomes to determine the link status of UE location $u$, (1) UE is present and unblocked, (2) UE is present but blocked, and (3) absent UE.
\begin{algorithm}
\caption{RIS-assisted blockage prediction}
\label{alg:algorithm_sum}
\begin{algorithmic}[1]
\State $\bold {Input}$: $\Gamma$, RIS panel assisted link and RGB image $X$ from ViWi dataset \cite{Ref_ViWi}
\State $\bold {Output}$: Maximizing the probability of the blockage prediction status
\State Generate $\cal B$ using the blockage prediction index elements, $i=\arg\max_{i \in \{1,2,...,U \}} \{ b_1,b_2,....,b_i,...b_U\}$
\State Calculate BS-to-RIS panel channel gain, $\mathbf{h_R}[k]$
\State Using Doppler spread, $f_s$, calculate BS-to-UE location $u$ and RIS-to-$u$ location channel gain, $\mathbf{h_B}[k]$ and $\mathbf{h_u}[k]$, respectively
\State Calculate $\log(\mathbf{I}+\mathbf{H_u^o}[k])$ using (\ref{OH})
\Repeat
\State Find optimal $Q_{\Gamma}(X,\log(\mathbf{I}+\mathbf{H_u^o}[k]))$ mapping to the blockage index, $\cal B$ with help of $(X,\log(\mathbf{I}+\mathbf{H_u^o}[k]))$ using ResNet-18 model
\State Use $X$ to detect the UE at location $u$, then move to $\log(\mathbf{I}+\mathbf{H_u^o}[k]))$ when $X$ is unable to detect UE
\State Determine if UE is absent or blocked, or unblocked
\State Using Step 9 and 10, train ResNet-18 model to maximize (\ref{Eq_opt})
\Until{convergence}
\end{algorithmic}
\end{algorithm}
\subsection{Solution}
The neural network model fed with an RGB camera and RIS panel-based channel modeling is presented in Fig.~\ref{System_Model2}.
We adopt residual network (ResNet)-18 neural network model, where 1) the model aims to learn and maximize $Q_{\Gamma}(X,\log(\mathbf{I}+\mathbf{H_u^o}[k]))$, 2) the model maximizes (\ref{Eq_opt}) using optimal $Q_{\Gamma}(X,\log(\mathbf{I}+\mathbf{H_u^o}[k]))$.
We train the model, ensuring it fits the blockage prediction model.
Note that $\Gamma$, and $(X,\log(\mathbf{I}+\mathbf{H_u^o}[k]))$ serve as the input of the neural network model.
This model is trained, followed by supervised learning, using the RGB image and RIS panel-assisted channel modeling from the environment labeling with their corresponding blockage index.
At least two neuron layers are fully connected in the model.
The adopted model is based on deep convolution neural networks, transferable learning, and predictive learning function.
Additionally, its final fully connected layer is removed and replaced with another fully connected layer with several neurons equal to the number of UE locations over $U$.
Additionally, $Q_{\Gamma }(X,\log(\mathbf{I}+\mathbf{H_u^o}[k]))$ maps
the RGB image and relates to a blockage probability, $b_i$.
During the training process, a cross-entropy loss is $o=\sum_{j=1}^U l_j \log b_j$, where $b_j$ is the probability mass function, which is induced by the soft-max layer.
The model also aims to find minimal cross-entropy loss
The relation between the cross-entropy loss nature and the UE location $u$ is inversely proportional, i.e., lower entropy loss results in finding the UE location $u$ tends to be deterministic than the random nature.
The detection of the UE at location $u$ is treated as a ternary classification problem, such as blocked, unblocked, or absent UE at location $u$.
The authors in \cite {MA_1} demonstrated that neural networks effectively learned to block prediction probability.
The network is, therefore, trained to detect the UE at location $u$, by using the blockage index, achieved from $Q_{\Gamma}(X,\log(\mathbf{I}+\mathbf{H_u^o}[k]))$.
At first, using $X$, the neural network model resolves of the UE is present and unblocked at location $u$.
If the detection at location $u$ is not successful, $\log(\mathbf{I}+\mathbf{H_u^o}[k]))$ detects the blockage scenarios, including present but blocked, and absent UE at location $u$.
This approach labels RIS panel-assisted link and image, $X$, and then uses them for the training model and is performed with ImageNet data \cite{OR_1}.
Using a deep neural network, blockage prediction involves detecting UE locations over a $U$, followed by RIS-assisted link assessment.
The detection of UE location $u$, therefore, maximizing the probability of blockage detection, is summarized in Algorithm 1.
 
Algorithm~\ref{alg:algorithm_sum} maximizes the probability of the blockage prediction.
$\cal B$ is produced in various UE locations $U$.
Using the RIS panel, $\mathbf{H_u^o}[k]$ is calculated.
Then ResNet-18 model plays a role in finding the existence of UE location $u$ using the dataset and results in the link status as $l_u=1$.
For an unsuccessful detection UE at the location, $u$, the RIS-assisted channel gain helps to detect whether $l_u$ is -1 or 0 or 1.

\section{Results}
In this section, we present the simulation results of the proposed model by discussing the dataset, neural network model, model training, code structure, and performance evaluation.

\textbf{Dataset:}
The following parameters are used to generate the dataset: the number of BS, UE, and MPCs are 1, 1, and 15, respectively.
Recall that the datasets for testing the blockage prediction solutions were generated using ViWi framework \cite{Ref_ViWi}.
ViWi provides various single-user communication scenarios and a Matlab script for generating synthetic dataset.
The dataset is composed of 5000 images. Using the generator package of ViWi, the corresponding images depicting the UE at some locations are generated.
 \begin{table}[]
\caption{Network fine-tuning parameters}
\label{Table_4}
\begin{center}
\begin{tabular}{ccc} \cline{1-3}
Batch size & 50 & 50 \\ \cline{1-3}
Learning rate & 1e-3 & 1e-3 \\ \cline{1-3}
Weight decay & 2e-3 & 2e-3 \\ \cline{1-3}
Learning rate schedule & epoch 5 and 8 & epoch 3 and 8 \\ \cline{1-3}
Learning rate reduction factor & 0.2 & 0.1 \\ \cline{1-3}
Data split & 70\%-30\% & 70\%-30\% \\ \cline{1-3}
Image graph & False & False \\ \cline{1-3}
\end{tabular}
\end{center}
\end{table}

\textbf{Neural network training:}
The parameters used to train the neural network model are shown in Table~\ref{Table_4}.
Each layer's normal distribution weight is initialized with a zero-mean and unit variance.
The network is fine-tuned based on the subset of the training data.
These parameters include the split between the datasets and are fine-tuned on the training subset of one of the two datasets.
In the deep learning model, there is a multilayer perceptron referred to as a feed-forward, which are universal function approximators.
This motivates adopting an MLP network to capture the relation between the environment descriptors and the RIS panel phase shift.

\textbf{Output of the training model:}
The parameters used for neural network training output are shown in Table~\ref{Table_2}.
We train our model with 18 hidden layers to vary the accuracy and reduce the entropy loss, which is minimized to 0.004613.
\begin{table}[]
\caption{Neural network training output}
\label{Table_2}
\begin{center}
\begin{tabular}{cc} \cline{1-2}
Name & Value \\ \cline{1-2}
Training size & 1 \\ \cline{1-2}
Output layer dimensions & 64 \\ \cline{1-2}
Epoch numbers & 10 \\ \cline{1-2}
Training batch numbers & 10-130 \\ \cline{1-2}
\end{tabular}
\end{center}
\end{table}

\begin{figure*}
\centering
\begin{subfigure}[b]{0.47\textwidth}
\centering
\includegraphics[width=\textwidth]{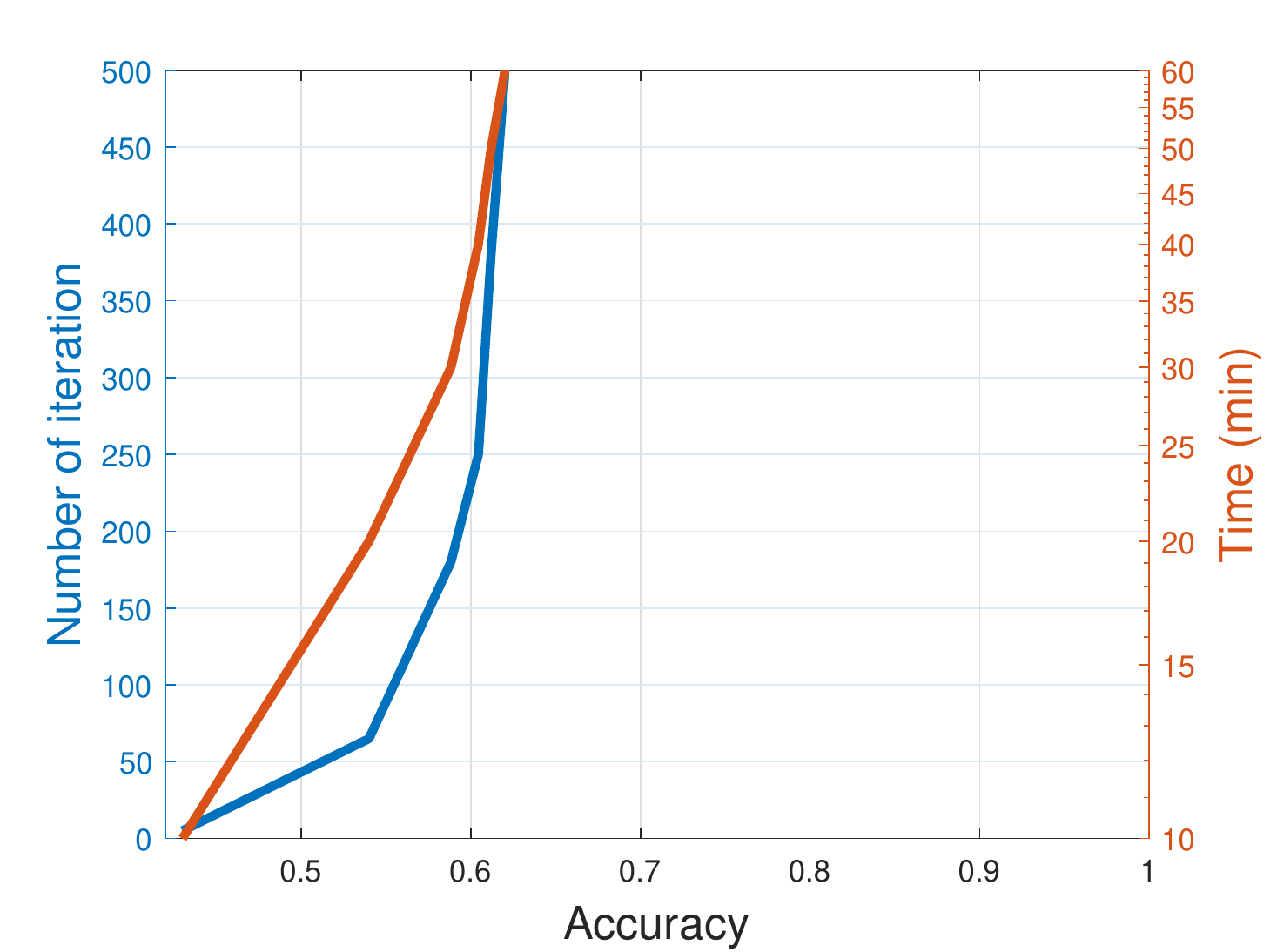}
\caption{No RIS panel and no RGB camera}
\label{P_a}
\end{subfigure}
\hfill
\begin{subfigure}[b]{0.48\textwidth}
\centering
\includegraphics[width=\textwidth]{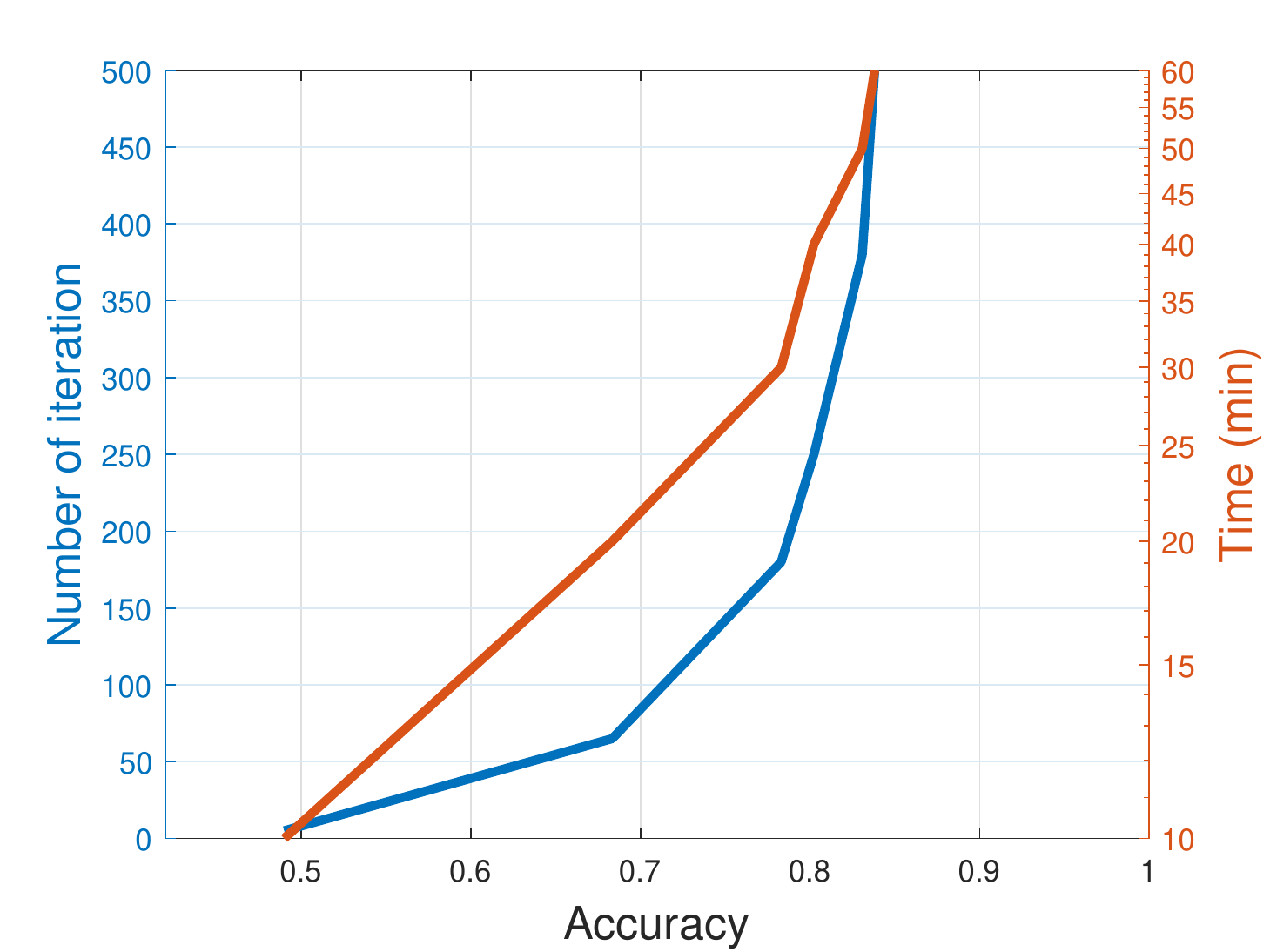}
\caption{RGB camera and no RIS panel}
\label{P_b}
\end{subfigure}
\hfill
\begin{subfigure}[b]{0.48\textwidth}
\centering
\includegraphics[width=\textwidth]{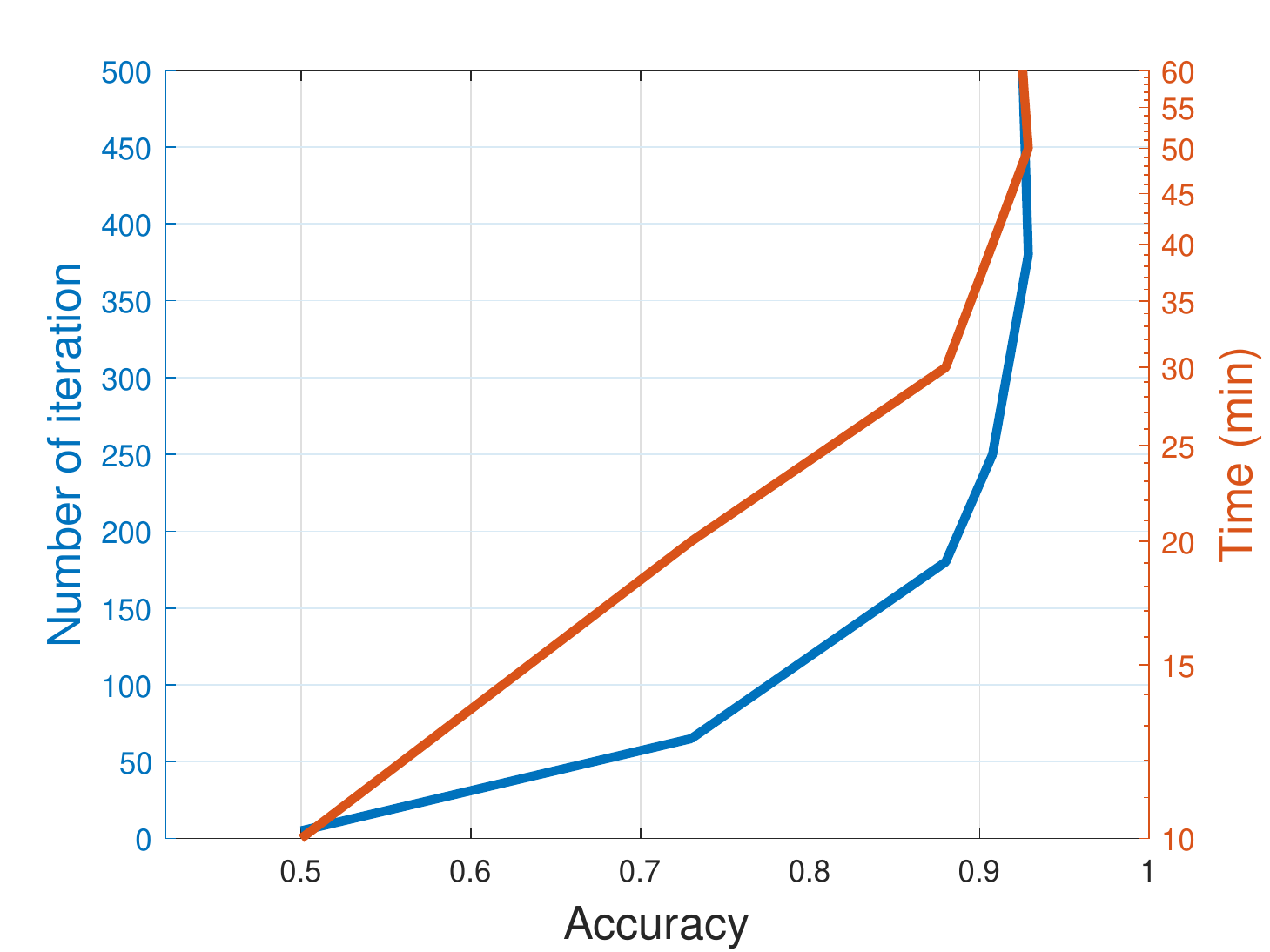}
\caption{RIS panel and no RGB camera}
\label{P_c}
\end{subfigure}
\hfill
\begin{subfigure}[b]{0.48\textwidth}
\centering
\includegraphics[width=\textwidth]{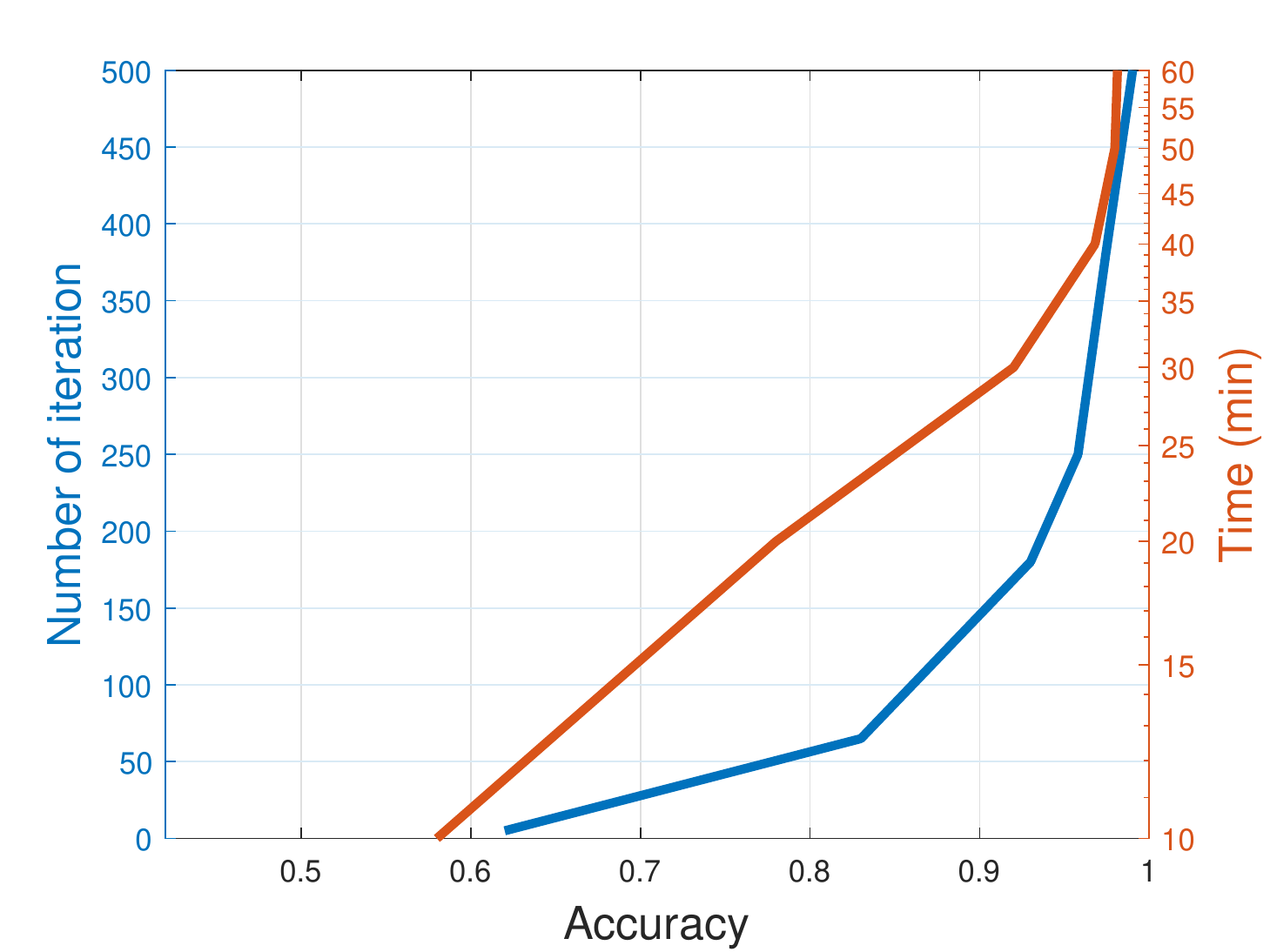}
\caption{RIS panel and RGB camera}
\label{P_d}
\end{subfigure}
\caption{Probability of blockage prediction accuracy varying with the number of iterations and time.}
\label{P}
\end{figure*}

We consider four different scenarios; hence, Fig.~\ref{P} shows the performance in terms of accuracy for those scenarios.
For example, the accuracy level is significantly low when the network has no RIS panel or RGB camera at the BS.
The accuracy level is around 61\% when 500 iterations are run over 60 minutes, as shown in Fig.~\ref{P_a}.
With an accuracy of around 83\%, Fig.~\ref{P_b} shows the model's performance when the network only adopts an RGB camera.
The accuracy raises 22\% compared to Fig.~\ref{P_a}.
The accuracy is obtained by considering the RIS panel in Fig.~\ref{P_c}.
The accuracy is estimated to be over 92\%.
The accuracy increases over 31\% compared to Fig.~\ref{P_a} and 9\% compared to Fig.~\ref{P_b}.
With an accuracy of over 99\%, Fig.~\ref{P_d} shows the model's performance when the networks adopt both the RIS panel and RGB camera.
The accuracy increases over 38\% compared to Fig.~\ref{P_a}, 16\% compared to Fig.~\ref{P_b}, and 7\% compared to Fig.~\ref{P_c}.
The proposed model has the best accuracy compared to Fig.~\ref{P_a}~-~Fig.~\ref{P_c}.
The proposed model has better accuracy over 38\% compared to the baseline scheme in Fig.~\ref{P_a}.
\section{Conclusion}
This paper introduces an approach for accurately predicting UE blockage using a combination of a BS equipped with an RGB camera and an RIS panel under practical consideration while considering mobility.
We adopt a ResNet 18-layer network and use a synthetic data set called "ViWi".
Considering MPCs, we compute the RIS panel-assisted channel gains between the BS to the RIS panel, the RIS panel to the UE, and the direct link between the BS and the UE.
We consider the UE mobility by incorporating the Doppler spread in the model to capture the effect of mobility in the system model.
Using RGB camera, ResNet-18 detects the presence without blockage of UE at various locations.
In case of an unsuccessful detection, the RIS-assisted link detects if the UE is either present but blocked or absent.
Therefore, we investigate the potential benefits of the RIS panel to enable wireless networks to help overcome the challenge of blocking prediction when the BS has RGB camera.
Finally, the extensive simulation results show the significantly improved accuracy of the proposed RIS panel-assisted framework for maximizing the probability of blockage prediction.

\end{document}